\documentclass[12pt]{article}
\pdfoutput=1
\usepackage{amsmath, amssymb}    
\usepackage{graphicx}
\usepackage{dcolumn}
\usepackage{bm}
\usepackage{hyperref}
\usepackage{color}

\newcommand{\be}{\begin{equation} } 
\newcommand{\ee}{\end{equation} } 
\newcommand{\ba}{\begin{array} } 
\newcommand{\ea}{\end{array} } 
\newcommand{\bear}{\begin{eqnarray} } 
\newcommand{\eear}{\end{eqnarray} }

\title{
\vspace*{-3.4cm}
\begin{flushright}
\normalsize{ \small   Fermilab-PUB-24-0807-T
  }
\end{flushright}
\vspace*{0.99cm}
\Large 
\textbf{TeV-scale particles and  LHC events with dijet pairs
 }\vspace*{0.3cm}   
}

\author{{\bf  \normalsize 
Bogdan A. Dobrescu} 
\vspace{5mm}
\\
\normalsize\emph{Particle Theory Department, Fermilab, Batavia, IL 60510, USA\footnote{bdob@fnal.gov}  
}
}
\date{ \normalsize  November 6, 2024 }

\begin{document}  
\setcounter{page}{0}  
\maketitle  

\begin{abstract}  
Scalar particles that couple to two up quarks may be produced at the LHC even if they are ultraheavy,  in the $8-10$ TeV mass range. A renormalizable theory that includes a diquark particle of this type ($S_{uu}$), two vectorlike quarks ($\chi_1, \chi_2$), and a gauge-singlet pseudoscalar,  predicts LHC signals  involving four or more jets of very high $p_T$. Two remarkable events observed by the CMS experiment, each involving four high $p_T$-jets, may be due to an $S_{uu}$ of mass near 8.5 TeV, and a $\chi_2$ mass of 2.1 TeV. A separate excess reported by CMS in the nonresonant dijet pair search is consistent with a $\chi_1$ mass of $0.95$ TeV. This hypothesis may be tested through CMS and ATLAS  searches for signals with a pair of dijet resonances of masses clustered around 1 TeV, and separately around 2 TeV, which have $4j$ invariant masses in the $5-8$ TeV range. These additional signals would arise from cascade decays of  $S_{uu}  \to  \chi_2 \chi_2 $, which  lead to $5j$ and $6j$ events with invariant masses around 8 TeV. Depending on the couplings of the heavy colored particles, additional signals are possible, involving for example  highly-boosted top quarks.

\end{abstract} 

\vfil

\thispagestyle{empty}  
  
\setcounter{page}{1}  
  
\vspace*{0.31cm}    
  
\newpage  
  
\tableofcontents
  
\vspace*{0.31cm}    
  
\baselineskip18pt   

\bigskip\bigskip

\section{Introduction} 
\label{sec:intro}

Searches for new physics at the LHC have become substantially more sensitive to new particles over the last decade \cite{ParticleDataGroup:2024cfk}.  
Particles that carry QCD color and have masses in the TeV range, in particular, are now probed at higher masses and lower couplings than ever before. This rapid progress is due to novel experimental methods, larger data sets, higher energies,  improved detectors, but also due to a wider set of signals being searched for.

An example of colored particle intensely searched for by the CMS and ATLAS experiments is a vectorlike quark ($\chi$) of electric charge 2/3.
Such a particle was proposed as a natural ingredient of theories that predict a composite Higgs boson \cite{Dobrescu:1997nm}. More generally,
vectorlike quarks are of great interest given that they would represent a markedly different type of fermion compared to the Standard Model (SM) ones, which are chiral.
Typically, a vectorlike quark like $\chi$ is expected to decay, through its mass mixing with the top quark, into $Wb$, $Z t$ and $h^0 t$ final states    \cite{Han:2003wu}. 
Searches  \cite{ATLAS:2018ziw}  for QCD production of  $\chi \overline\chi$ followed by the above decay modes have set lower mass limits of about 1.4 TeV. It is possible, though, that the mass mixing of $\chi$ is suppressed such that the vectorlike quark predominantly decays to other final states \cite{Dobrescu:2016pda}, for example through an off-shell diquark particle.

It was proposed in Ref.~\cite{Dobrescu:2018psr} that the leading decay of $\chi$ may be into an up quark and a gluon, through a dimension-5 operator.
Furthermore, pair production of $\chi$ may proceed through an $s$-channel 
hypothetical particle, $S_{uu}$, which is a diquark scalar that transforms under the QCD gauge group as a color sextet, and carries electric charge 4/3. 
The LHC signature in that case is a pair of dijet resonances, each of them peaked at the mass ($m_\chi$) of $\chi$, with the invariant mass 
of the leading four jets near the mass $M_S$ of $S_{uu}$. A peculiar event reported earlier by CMS \cite{CMS:2018wxx} with two wide jets forming an invariant mass of 8 TeV, each of the wide jets having an invariant mass in the $1.8-1.9$ TeV range and a definite two-prong substructure, could be produced by a  $pp \to S_{uu} \to \chi\chi \to 4j$ process \cite{Dobrescu:2018psr}. The QCD background for such a high-mass event was estimated in \cite{Dobrescu:2018psr} to be below $10^{-4}$ events. 
An alternative interpretation was proposed in \cite{Dobrescu:2019nys}, namely the decay of $S_{uu}$ into two diquarks of charge 2/3, each of them subsequently  decaying into two SM antiquarks. Other LHC signatures of $S_{uu}$ are also explored in \cite{Dobrescu:2019nys}.

In 2022, a dedicated search by CMS \cite{CMS:2022usq} for the $pp \to S_{uu} \to \chi\chi \to 4j$ process yielded a $3.9\sigma$ excess over the QCD background, due to two events (the earlier event plus a new one) consistent with $M_S \approx 8.5$ TeV and $m_\chi \approx 2.1$ TeV.  
ATLAS \cite{ATLAS:2023ssk} performed a similar search, and found no events with $4j$ invariant mass above 7 TeV. However, ATLAS observed one outlier event with the 4-jet invariant mass $M_{4j} \approx 6.6$ TeV and the average dijet mass $M_{2j} \approx 2.2$ TeV. That event may still be produced by an $S_{uu}$ because not all QCD radiation is captured in the jet cones, and very high-mass resonances have a long tail \cite{Harris:2011bh} towards lower masses where the PDFs are exponentially larger.
The same CMS article \cite{CMS:2022usq} presented a search for nonresonant production of a pair of dijet resonances, modeled as squark pair production, and found a $3.6\sigma$ excess at $M_{2j} \approx 0.95$ TeV. The production rate consistent with that excess is larger than the one for squarks by an order of magnitude.

The present article points out that QCD production of $\chi\overline\chi$ followed by vectorlike quark decays into two jets has the correct cross section for explaining the  CMS nonresonant $4j$ excess if the $\chi$ mass is 0.95 TeV. A renormalizable theory that includes a gauge-singlet pseudoscalar ($A_\chi$) coupled to $\chi$ and to the up quark gives rise to a dijet decay of  $\chi$ for two mass ranges of $A_\chi$. The first one is $M_A > m_\chi$, which leads to
$\chi \to u g $ at one loop. The second  mass range is $M_A \ll m_\chi$, which results in  $\chi \to u A_\chi  $ at tree level, followed by $A_\chi \to gg $ at one loop; the boost of $A_\chi$ is very high in this mass range, so the two gluons are nearly collinear and form a single narrow jet.

An extended renormalizable theory, also presented here, which includes an $S_{uu}$ diquark, two vectorlike quarks ($\chi_1$, $\chi_2$) and an $A_\chi$ pseudoscalar, offers a consistent and simultaneous explanation of the CMS  resonant and  nonresonant  $4j$ excess events. Moreover, this theory predicts cascade decays of $S_{uu}$ that lead to events similar to the ATLAS outlier event \cite{ATLAS:2023ssk}. 

Section~\ref{sec:1tev} focuses on the predictions of the theory that includes, beyond the SM, only a $\chi$ and an $A_\chi$.  
In Section~\ref{sec:minimal}, the theory that also includes the $S_{uu}$ diquark is studied, and the invariant mass distribution of the four jets is computed for new particle masses consistent with the CMS 8 TeV events.
Related LHC signatures are also discussed there.
The extended renormalizable theory with two vectorlike quarks is proposed and analyzed in Section~\ref{sec:2Vquarks}. 
The conclusions are summarized in Section~\ref{sec:conclusions}. 

\medskip  

\section{A vectorlike quark decaying into two jets}     
\label{sec:1tev}   \setcounter{equation}{0}

Consider a vectorlike quark $\chi$ of charges $(3,1,+2/3)$ under the $SU(3)_c \times SU(2)_W \times U(1)_Y$  gauge group, whose  chiral components ($\chi_L$ and $\chi_R$) transform under 
an approximate symmetry, {\it e.g.} a global  $U(1)$, such that the mass mixings of $\chi$ with the top, charm or up quarks are negligible.
Assume that the global symmetry is broken by the interactions of a gauge-singlet pseudoscalar $A_\chi$  with $\chi$:  
\be
- i A_\chi  \,  \overline \chi_L \left(  \lambda_\chi \,  \chi_R +  \lambda_u  \, u_R \right)  + {\rm H.c.}  ~~,
\label{eq:yuk1}
\ee
where  $\lambda_\chi, \lambda_u$ are Yukawa couplings. Similar couplings to the top or charm quarks are possible, but here they are assumed to have  Yukawa couplings much smaller than $\lambda_u$.

\subsection{Large $\chi \to jj$ branching fraction}

For a mass $m_\chi$ of $\chi$ at the TeV scale, there are two ranges for the mass $M_A$ of $A_\chi$  that lead to a decay of  $\chi$ into two jets with a branching fraction close to 100\%.

\subsubsection{Light-$A_\chi$ case: $M_A \ll m_\chi$}  
\label{sec:light}

\begin{figure}[t!]
\vspace*{6mm}  
\hspace*{0.5cm}
 \includegraphics[width=6.5cm, angle=0]{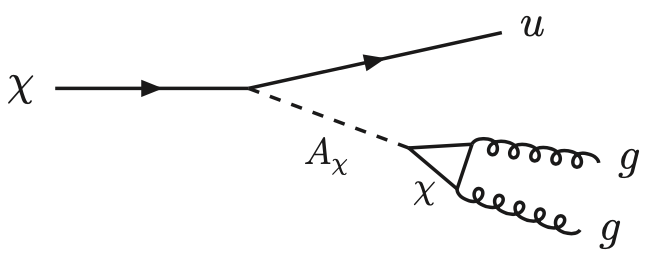} \\ [-3.1cm]
\hspace*{8.7cm}
\includegraphics[width=5.9cm, angle=0]{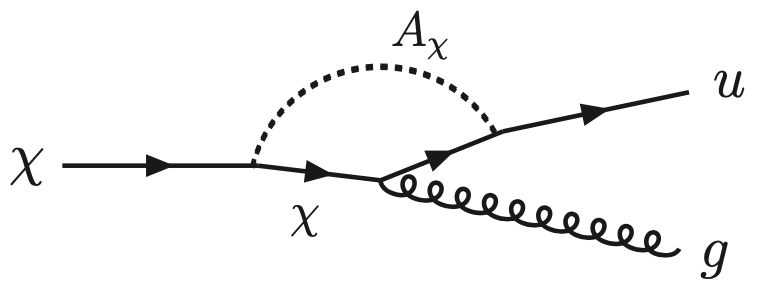} 
\vspace*{.5cm}  
\caption{Vectorlike quark $\chi$ decays into two jets. Left diagram occurs in the $M_A \ll m_\chi$ case, and includes a highly boosted $A_\chi$ pseudoscalar that gives rise to a $j_{gg}$  
jet initiated by two collinear gluons; this leads to the $\chi \to j \, j_{gg}$ process.
Right diagram corresponds to the $M_A > m_\chi$ case, and represents the loop-induced decay  $\chi \to u \, g$.
\vspace*{2mm}
} 
\label{fig:chiDecay}
\end{figure}

When $A_\chi$ is light, the only tree-level decay of $\chi$ is $\chi \to u \,   A_\chi $ with a highly boosted $A_\chi$. Subsequently,  $A_\chi$ decays through a $\chi$ loop into two gluons. Due to the boost of  $A_\chi$,  the $A_\chi \to gg $ signal appears in the detector as a single narrow jet, with substructure, labelled here $j_{gg}$. Thus, $\chi$ effectively decays into two jets: $\chi \to j \,  j_{gg} $, where $j$ is the hadronic jet initiated by the up quark. 
This process is shown in the first diagram of Figure~\ref{fig:chiDecay}.  Additional decay modes of  $A_\chi$, such as  $A_\chi \to \gamma\gamma $ have suppressed branching fractions.

A wide range of values for the $\lambda_u$ coupling in (\ref{eq:yuk1}) is allowed.
The decay length of $\chi$ is proportional to $\lambda_u^{-2}/m_\chi $, such that $\chi$ decays promptly inside an LHC detector as long as the lower limit $\lambda_u \gtrsim O(10^{-6})$ is satisfied. An upper limit $\lambda_u \lesssim O(10^{-2})$ is necessary to suppress too large $\chi A_\chi$  production at the LHC from an initial state $u$. 

Constraints on $ \lambda_\chi$ are more stringent. 
The decay length of $A_\chi$ is proportional to $ \lambda_\chi^{-2}  m_\chi^2/M_A^3 $; this loop-suppressed decay is prompt provided $\lambda_\chi \gtrsim 5 \times 10^{-3}$ for $M_A \approx 10 $ GeV \cite{Bernreuther:2023uxh}. 
 An upper limit $\lambda_\chi \lesssim 0.1$ is necessary to suppress too large $A_\chi$  production at 1-loop through gluon fusion. Note though that the coupling  limit on an $s$-channel  dijet resonance of mass in the $5-100$ GeV range depends in a complicate way on the resonance mass \cite{Dobrescu:2021vak} (because $A_\chi$ has spin 0,  it does not mix with the $Z$ or the $\Upsilon$ meson, so the limits are mostly due to LHC searches).

\subsubsection{Heavy-$A_\chi$ case:  $M_A > m_\chi$}  
\label{sec:heavy}

When $A_\chi$ is heavy,  $\chi$ is stable at tree level but decays at one loop, through the second diagram of Figure~\ref{fig:chiDecay}. The main decay mode is 
$\chi \to u g$, which again is a decay into two jets with a branching fraction near 100\%.
Other 1-loop decay modes exist (see \cite{Kim:2018mks} for a similar discussion), including $\chi \to u \gamma$ and $\chi \to u Z$, but with rates suppressed by $\alpha/(3 \alpha_s)$. In addition,  a loop involving the $A_\chi$ pseudoscalar and  $\chi$  induces a mass mixing of $\chi $ and $u$, which leads to the  $\chi \to d \, W,  \, u \, Z, \, u \, h^0$ decay modes.

Integrating out the heavy  $A_\chi$ leads to a dimension-5 operator  \cite{Kim:2018mks} that, at energy scales $E < M_A$, approximates (with corrections of order  $m_\chi^2/M_A^2$ and $E^2/M_A^2$) the interaction of the gluon field strength $G^a_{\mu \nu}$ with the vectorlike quark and the SM right-handed $u$ quark:
\be
\frac{1}{\mu_G}  \,  G^a_{\mu \nu} \, \overline u_R  \, \gamma^\mu \gamma^\nu T^a \chi_L + {\rm H.c.}  ~,
\label{eq:dim5}
\ee
where $T^a$ is an $SU(3)_c$ generator, and $\mu_G$ is a mass parameter given by
\be
\mu_G = \frac{16\pi^2 \, M_A^2 } { \lambda_\chi  \lambda_u  \,  m_\chi  \ln (M_A/m_\chi) }  ~~.
\label{eq:coef-dim5}
\ee
Since $\lambda_\chi  , \lambda_u  \lesssim O(1)$,  $m_\chi \ll \mu_G$  so $\chi$ is a very narrow dijet resonance. 
As long as $m_\chi/\mu_G \gtrsim O(10^{-6}) $, $\chi$ decays promptly inside an LHC detector. 
Note that the effective interaction (\ref{eq:dim5}) could also be induced by heavy fields other than $A_\chi$, in which case the expression 
(\ref{eq:coef-dim5}) for $\mu_G$ would be modified. 

\bigskip

\subsection{$\chi \overline{\chi} $ production at the LHC}

\begin{figure}[t!]
\hspace*{2.5cm}  \includegraphics[width=10.2cm, angle=0]{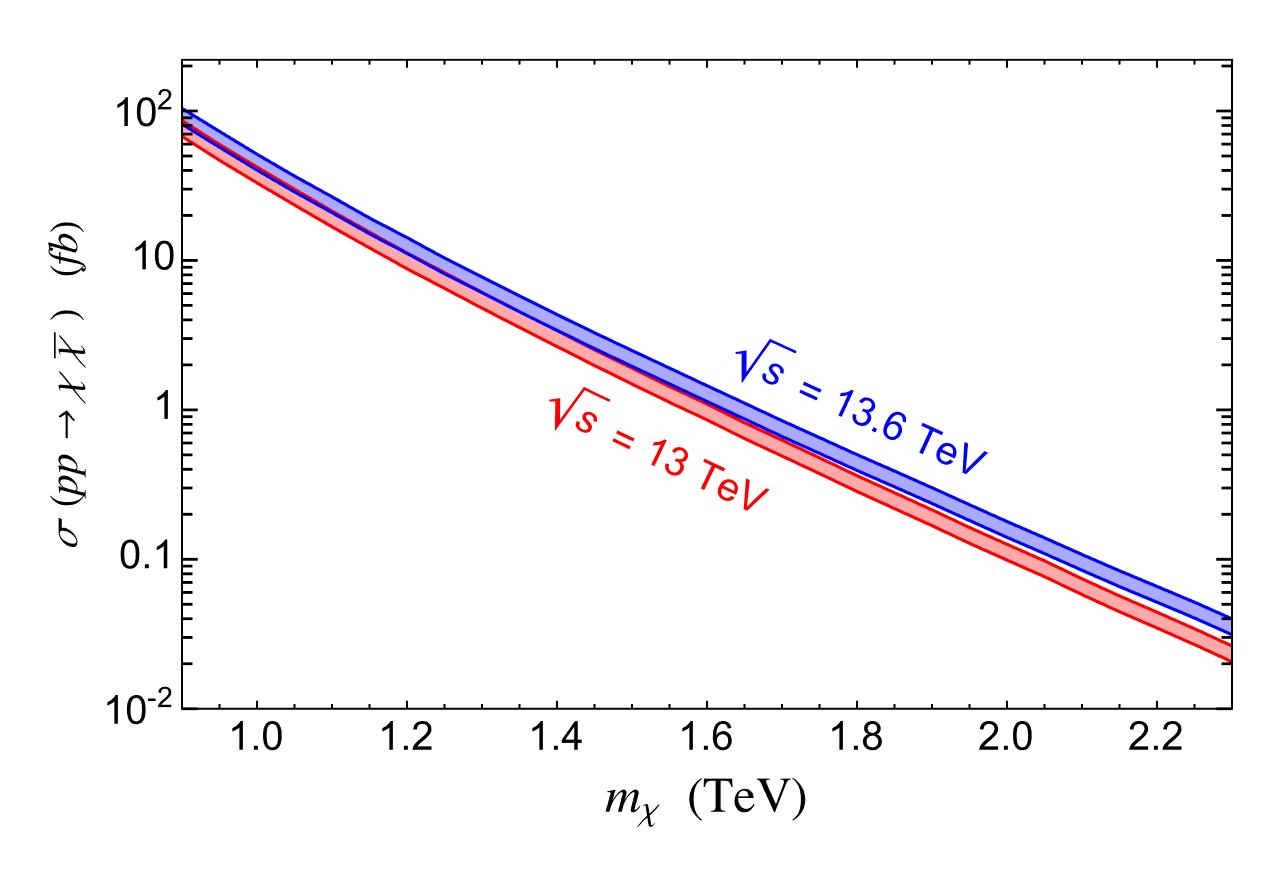}     
\vspace*{-.2cm}  
\caption{LHC cross section for QCD production of $\chi \overline\chi$, as a function of the  vectorlike quark mass $m_\chi$. Lower (red) and upper (blue) bands correspond to $pp$ center-of-mass energies of 13 TeV and 13.6 TeV, respectively.  The cross section is computed with Madgraph \cite{Alwall:2014hca} at NLO, and the band widths indicate uncertainties obtained by changing the renormalization and factorization scales  by a factor of 1/2 or 2.
\vspace*{2mm}
} 
\label{fig:xsecChi}
\end{figure}

The main production mode of $\chi$ at the LHC is $\chi \overline\chi$, through its QCD coupling to the gluon. This process is the same as the top quark pair production in QCD, except for $m_\chi \gg m_t$.
Figure~\ref{fig:xsecChi} shows the production cross section of $\chi \overline\chi$ at the LHC as a function of the only free parameter, $m_\chi$. 
That cross section is computed at next-to-leading order (NLO) with Madgraph \cite{Alwall:2014hca}, version MG5\_aMC\_v3\_5\_6 and the PDF set
NNPDF23\_nlo\_as\_0119\_qed  \cite{Ball:2013hta}.
The lower red band is for $\sqrt{s} = 13$ TeV (LHC Run 2), and the upper blue  band is for $\sqrt{s} = 13.6$ TeV (current Run 3 of the LHC).
The width of each band represents the uncertainties given by changes in the renormalization and factorization scales by a factor of 1/2 or 2.

A representative diagram for the production of $\chi \overline\chi$ followed by the  decays of the vectorlike quarks 
into $u A_\chi$ (left diagram, for the Light-$A_\chi$ case) or $ug$ 
(right diagram, for the Heavy-$A_\chi$ case) is shown in Figure~\ref{fig:DiagramsNonreschi}.
The signal is a pair of dijets, with the two dijets having the same invariant mass given by $m_\chi$.
Searches for signals of this type have been performed by both CMS \cite{CMS:2022usq, Sirunyan:2018rlj} and ATLAS \cite{Aaboud:2017nmi}.

\subsection{CMS excess of dijet-pair events at $m_{jj} = 0.95$ TeV}
\label{sec:950GeV}

The most sensitive search to date has been performed by CMS (the nonresonant search in \cite{CMS:2022usq}) with 138 fb$^{-1}$ of 13 TeV data. The CMS result shows an excess of events for a dijet mass $m_{jj} = 0.95$ TeV, with a local significance of 3.6$\sigma$.
The observed 95\% CL limit on the cross section ($\sigma$) times the product of branching fractions (${\cal B}_{jj}$) and acceptance ($A_{4j}$) 
 is approximately $\sigma  \, {\cal B}_{jj} \, A_{4j} < 8 $ fb at $m_{jj} = 0.95$ TeV, while the expected limit was $3 $ fb. 

\begin{figure}[t!]
\vspace*{-0.1cm} 
\hspace*{-0.1cm}  \includegraphics[width=0.52\textwidth]{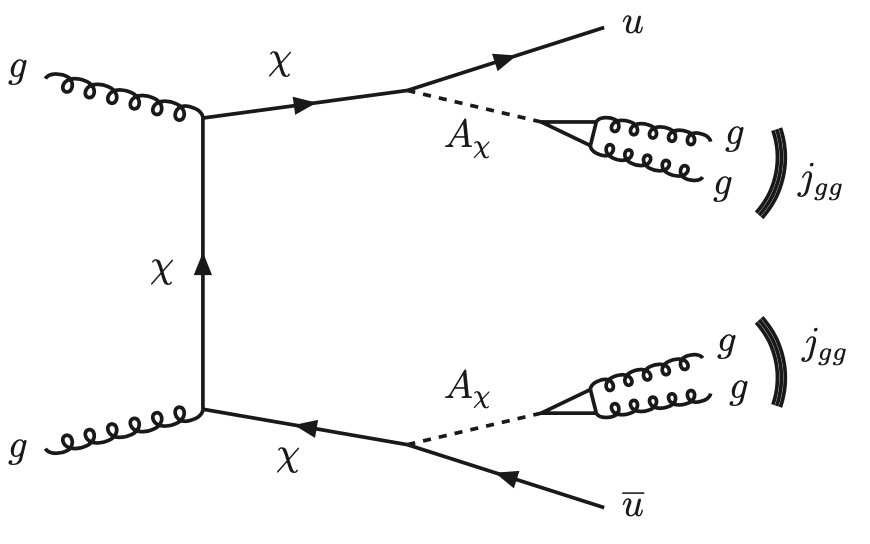}
 \\ [-4.9cm]
\hspace*{9.5cm}  \includegraphics[width=0.39\textwidth]{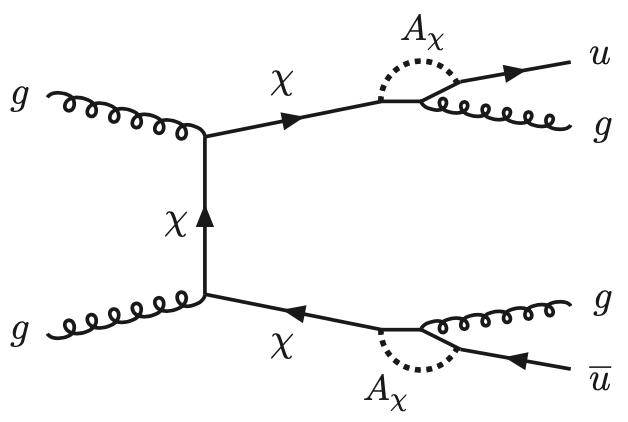}
\vspace*{4mm}
\caption{LHC production of  $\chi \overline \chi$ with a pair of dijets in the final state. 
Left diagram ($M_A \ll m_\chi$): vectorlike quark decay into $u\, A_\chi$ followed by the loop decay of $A_\chi$  into two collinear gluons which form a single jet, $j_{gg}$.
Right diagram ($M_A > m_\chi$): vectorlike quark decay into $u\,g$ through a 1-loop process involving virtual $A_\chi$ and $\chi$ fields.
\vspace*{3mm}
} 
\label{fig:DiagramsNonreschi}
\end{figure}

To compute the acceptance in the case of the $pp \to \chi \overline\chi \to 4j$ process, the FeynRules package \cite{Alloul:2013bka} was used to generate Madgraph \cite{Alwall:2014hca} model files that include the SM plus a color-triplet $\chi$ fermion and the following: $i$) an $A_\chi$ with interactions (\ref{eq:yuk1}) for the light-$A_\chi$ case (default mass $M_A = 10$ GeV), or 
$ii$) just the dimension-5 interaction (\ref{eq:dim5}) for the Heavy-$A_\chi$ case.

The Madgraph simulation at parton level and at leading order in the QCD coupling (with the default PDF set used in Madgraph, NNPDF23\_lo \cite{Ball:2013hta}), followed by the event selection described in \cite{CMS:2022usq}, yields an acceptance $A_{4j}$ as a function of $m_\chi$.
The pairing of the two jets in each dijet is decided within each event by the following CMS algorithm: for each of the 3 possible pairings of the leading four jets,
the pairing that minimizes the function $|\Delta R_1 - 0.8|^2 + |\Delta R_2 - 0.8|^2$ is selected, where $\Delta R_1$ is the angular separation of the two jets inside the first dijet, and $\Delta R_2$ is the angular separation inside the second dijet.
After the pairing is  chosen, strong cuts of $\Delta R_1 < 2$ and $\Delta R_2< 2$ are imposed to select events where the two dijets are boosted, in order to suppress the QCD background. Another cut, on the $\eta$ separation between the two dijets, $\left|\Delta\eta_{12}\right| < 1.1$ is also imposed. Finally, there is a cut on the dijet mass asymmetry:  $|m_1 - m_2|/(m_1 + m_2) < 0.1$ where $m_1$ and $m_2$  are the invariant masses of each dijet.
For $m_\chi = 0.95$ TeV, the computed acceptance of the above kinematic selection criteria is $A_{4j} = 8.3\%$ for the Light-$A_\chi$ model, and 
$A_{4j} = 8.0\%$ for the Heavy-$A_\chi$ model.

The cross section for $\chi\overline{\chi}$ production at the 13 TeV LHC is $\sigma(\chi\overline{\chi}) \approx 54$ fb at  $m_\chi = 0.95$ TeV. As the $\chi \to jj$ branching fraction is ${\cal B}(\chi \to jj) \approx 100\%$, the combined  branching fraction is ${\cal B}_{jj} = {\cal B}(\chi \to jj)^2 \approx 1$, which gives $\sigma(\chi\overline{\chi}) \, {\cal B} _{jj} \, A_{4j}  \approx  4.3$ fb. This is between 
the expected limit ($\sim 3$ fb)  and the observed limit ($\sim 8$ fb). 
Thus, the process of Figure~\ref{fig:DiagramsNonreschi} with $m_\chi = 0.95$ TeV  has a rate consistent with the excess in the  CMS nonresonant search  \cite{CMS:2022usq}, even though there are no adjustable parameters (other than $m_\chi$) in this theory.\footnote{An alternative explanation \cite{Crivellin:2022nms} is based on a diquark decay into a pair of other diquarks, similar to the process proposed in  \cite{Dobrescu:2019nys} but at lower masses. Another explanation discussed in \cite{Crivellin:2022nms}, involving colorons of 1 TeV,  appears to be in conflict with the large QCD production cross section for a coloron pair \cite{Dobrescu:2007yp}.}
The cross section for $\chi\overline{\chi}$ production increases by approximately 22\% at the 13.6 TeV LHC:  $\sigma(\chi\overline{\chi}) \approx 66$ fb. Hence, if a $\chi$ with $m_\chi \approx 0.95$ TeV and  ${\cal B}(\chi \to jj) \approx 100\%$ exists, then a $5\sigma$ discovery in Run 3 is possible with less than 300 fb$^{-1}$.

\begin{figure}[t!]
 \hspace*{2.4cm}  \includegraphics[width=0.68\textwidth]{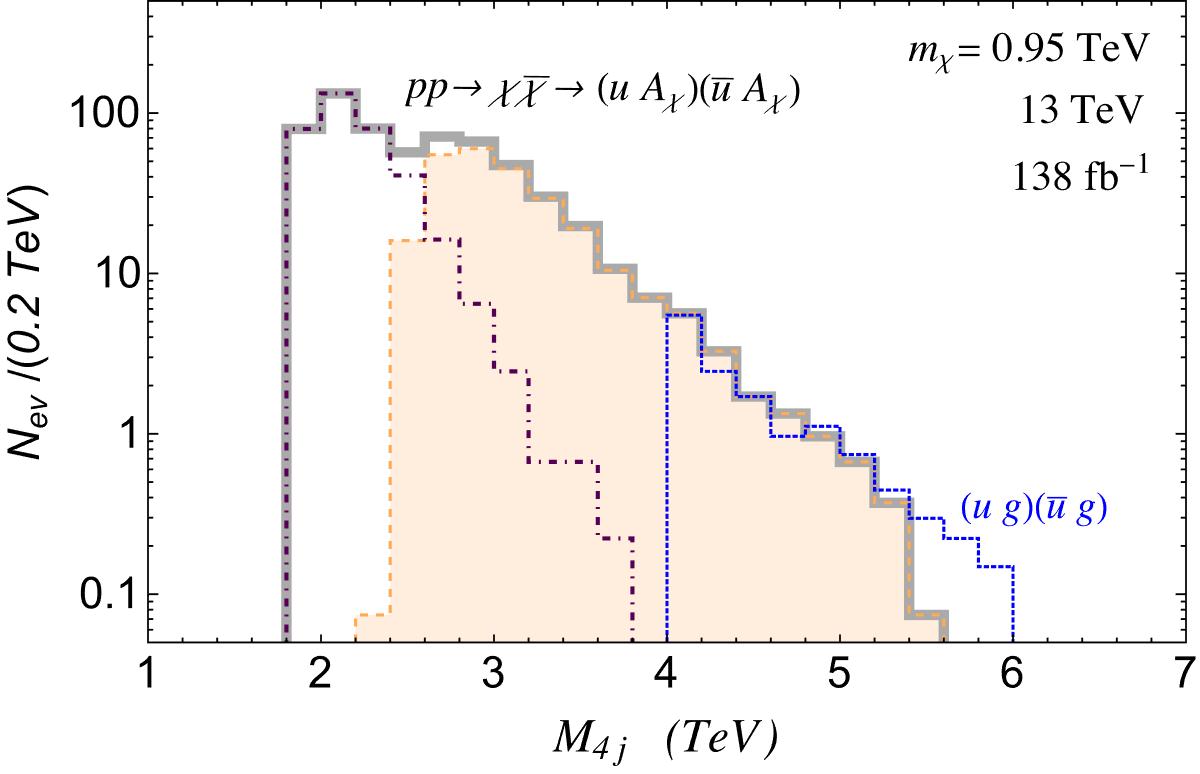}  
 \\ [-0.6cm]
\caption{Invariant mass distribution of the 4 jets produced at 
$\sqrt{s} = 13$ TeV in the nonresonant process 
$pp \to \chi \overline \chi \to   (u A_\chi)( \overline u A_\chi ) \to 4j$ simulated  at parton level for $m_\chi = 0.95$ TeV and $M_A = 10$ GeV. 
This simulation uses Madgraph at leading order,  
integrated luminosity of 138 fb$^{-1}$, and event selection described in the text.
Dashed orange line (boundary of the shaded region) represents the distribution of events that have the correct pairing of the two dijets (based on their $\chi$ or $\overline \chi$ origin). Dot-dashed purple line represents the distribution of events with incorrectly paired dijets that still satisfy the selection. Solid thick gray line is the sum of these two distributions. For comparison, the dotted blue line shows the total $M_{4j}$ distribution for $pp \to \chi \overline \chi \to (u g)( \overline u g)$ in the Heavy-$A_\chi$ case, only for $M_{4j} > 4$ TeV (at lower $M_{4j}$ it is almost the same as the solid thick line).
 \vspace*{2mm}
} 
\label{fig:plotNonres4j1000}
\end{figure}

The $4j$ invariant mass distribution, obtained by simulating the $pp \to \chi \overline \chi \to   (u A_\chi)( \overline u A_\chi )$ process at parton level with 138 fb$^{-1}$ of LHC collisions at $\sqrt{s} = 13$ TeV using the selection criteria discussed above, is shown as the solid thick gray line in Figure~\ref{fig:plotNonres4j1000}. The shaded  region  (bounded by the dashed orange line) shows the number of events (per bins of 0.2 TeV) in which the dijets are correctly paired, while in the remaining events (with $M_{4j}$ distribution given by the dot-dashed purple line) each dijet includes a jet ($u$ or $j_{gg}$) from $\chi$ and one from $\overline \chi$. The $M_{4j}$ distribution of correctly paired events (which has an average $2j$ invariant mass distribution sharply peaked at  $m_\chi$) extends to large  invariant masses (there are 1.1 events expected with $M_{4j} > 5$ TeV). By contrast, $M_{4j}$ is shifted to lower values in the case of incorrectly paired jets.

For comparison to the Heavy-$A_\chi$ case, the dotted blue line gives the total $M_{4j}$ distribution for $pp \to \chi \overline \chi \to (u g)( \overline u g)$; that is displayed only for $M_{4j} > 4$ TeV because at lower $M_{4j}$ the distribution is almost indistinguishable from the thick gray line.
As can be seen in Figure~\ref{fig:plotNonres4j1000}, the Heavy-$A_\chi$ case has a distribution with a longer tail at large $M_{4j}$: there are 1.9 events expected with $M_{4j} > 5$ TeV. Differences of this type between the two processes are due to the different spins of the emitted particles (spin-0 $A_\chi$ versus spin-1 gluon), which lead to changes in the angular correlations.
Parton shower, hadronization, jet algorithms and detector effects would further  broaden all the distributions. 


\section{$S\chi A$ theory for ultraheavy dijet pairs}        
\label{sec:minimal}   \setcounter{equation}{0}

Consider now a renormalizable theory (referred to as $S\chi A$) that, in addition to the vectorlike quark $\chi$ and the pseudoscalar $A_\chi$ introduced at the beginning of  Section \ref{sec:1tev}  [see Lagrangian terms (\ref{eq:yuk1})], 
includes a scalar diquark $S_{uu}$ that transforms under the $SU(3)_c \times SU(2)_W \times U(1)_Y$ 
gauge group as $(6,1, +4/3)$.   
These gauge charges allow the following Yukawa interactions  in the Lagrangian:
\be
\dfrac{1}{2} \; K^n_{ij} \; S_{uu}^n \;  \left(   y_{uu} \,  \overline u_{R \, i} \,  u^c_{R \, j}
+ y_{\chi\chi_R} \, \overline \chi_{R \, i} \,  \chi^c_{R \, j}   +  y_{\chi\chi_L} \, \overline \chi_{L \, i} \,  \chi^c_{L \, j} 
 \, \right)  
+ {\rm H.c.}  ~~,
\label{eq:diquarkChi}
\ee
where  $y_{uu}$, $y_{\chi\chi_R}$ and $y_{\chi\chi_L}$ are dimensionless  parameters.
The upper index $n$ labels the sextet color states ($n=1,..., 6$), the $i$, $j$ indices label the triplet color states ($i,j = 1,2,3$), and the coefficients $K^n_{ij}$ arise from products of $SU(3)_c$ generators \cite{Luhn:2007yr}:
\be
K^{2\ell + 1}_{ij} = \delta_{i \ell} \delta_{j \ell}      \;\;\; \;\;  ,    \;\;\; \;\; 
K^{2\ell }_{ij} = \dfrac{1}{\sqrt{2} } \left(  \delta_{i \ell } \delta_{j \ell'} +  \delta_{i \ell'} \delta_{j \ell}   \right)    \;\;\; \;\;  ,
\label{eq:Kns}
\ee
where $\ell = 1,2,3$, and $\ell' = (\ell +1) \; {\rm mod} \, 3$.  

Additional Yukawa interactions of $S_{uu}$, for example to $\overline t_{R } \,  t^c_{R }$,
$\overline t_{R } \,  u^c_{R }$, $\overline \chi_{R } \,  u^c_{R }$, $\overline \chi_{R } \,  t^c_{R }$, may exist (see \cite{Dobrescu:2019nys} for a discussion of their phenomenological implications),  but  for simplicity  are assumed here to be negligible. The complex phases of $y_{uu}$ and  $y_{\chi\chi_R}$ may be absorbed in the $S_{uu}$ and $\chi$ fields, but $y_{\chi\chi_L}$ is generically complex.

\subsection{Decay widths and cross sections}

The mass of $S_{uu} $, $M_S$, is taken to satisfy $M_S > 2 m_\chi$.
Two tree-level decay modes are allowed by (\ref{eq:diquarkChi}), $S_{uu} \to uu, \chi\chi $, with leading-order partial widths   
\bear
 && \Gamma (S_{uu} \to uu) =  \dfrac{ y_{uu}^2 }{32 \pi}  M_S ~~,
\nonumber \\ [2mm]
 && \Gamma (S_{uu} \to \chi \, \chi)  =   \dfrac{ y_{\chi}^2 }{32 \pi}  M_S   \;   \left( 1 - \dfrac{2 m_\chi^2 }{M_{S}^2 } \right)\left( 1 - \dfrac{4 m_\chi^2 }{M_{S}^2 } \right)^{\! 1/2}  ~~,
\label{eq:uuWidth}
\nonumber
\eear
where $y_{\chi} $ is an effective coupling,
\be   
y_{\chi} = \left( y_{\chi\chi_R}^2 + |y_{\chi\chi_L}|^2 \right)^{1/2} ~~.  
\label{eq:ychi}
\ee
Note that the interactions (\ref{eq:diquarkChi}) allow $S_{uu}$ decays into two $\chi$'s, and not into $\chi \overline \chi$.

\begin{figure}[t!]
\hspace*{-0.17cm}  
\includegraphics[width=0.52\textwidth]{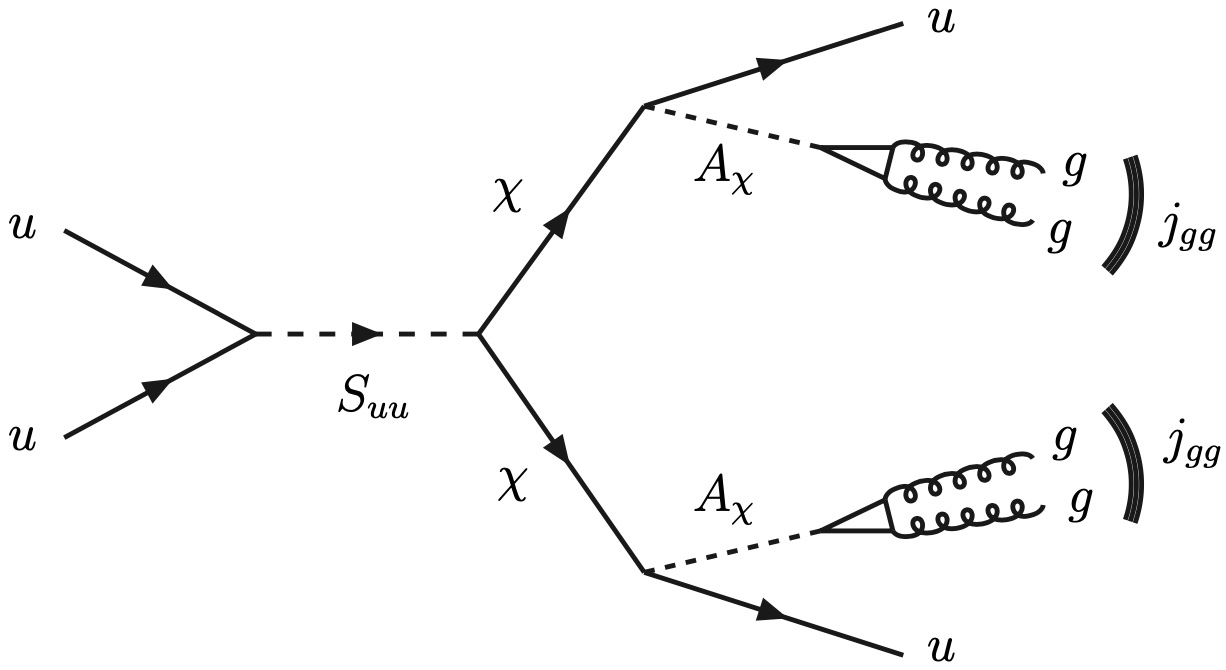} \\ [-3.95cm]
\hspace*{9cm}  \includegraphics[width=0.425\textwidth]{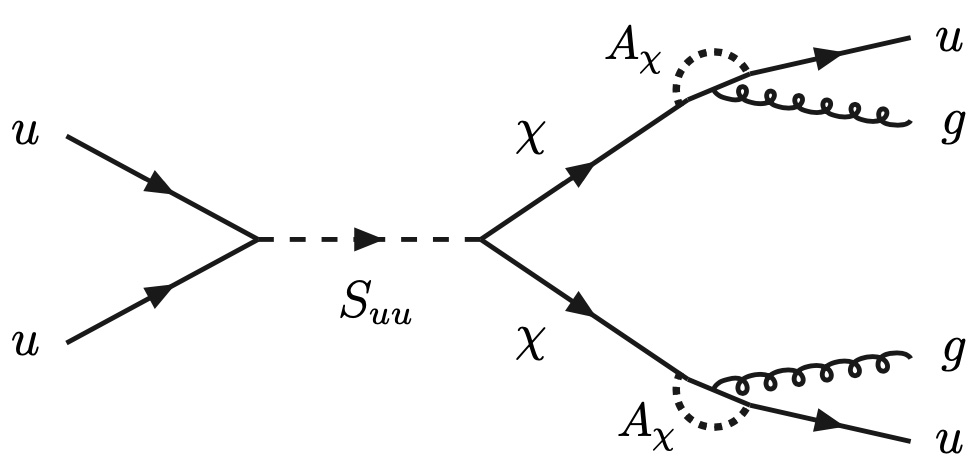} 
 \\ [0.1cm]
\caption{$s$-channel production of the $S_{uu}$ diquark at the LHC  followed by a cascade decay via $\chi\chi$ leading to a dijet pair. 
Left diagram ($M_A \ll m_\chi$): each $\chi$  decays into $u\, A_\chi$, with the pseudoscalar $A_\chi \to gg $ forming a single $j_{gg}$ jet.
Right diagram ($M_A  >  m_\chi$): each $\chi$  decays into $u\,g$ through a 1-loop process involving virtual $A_\chi$  and  $\chi$ fields.
\vspace*{2mm}
} 
\label{fig:DiagramSchichi}
\end{figure}

At the LHC, $S_{uu} $ may be produced in the $s$ channel due to its couplings to two up quarks, which have the largest PDFs for $M_S$ in the TeV range.  The cross section for its production was computed in \cite{Dobrescu:2018psr} as a function of $M_S$ based on the NLO analytical formulas from \cite{Han:2009ya}. After production, $S_{uu} $ decays into $uu$ giving rise to a dijet resonance  peaked at $M_S$, or it 
undergoes a cascade decay that depends on the pseudoscalar mass $M_A$: 
\bear
&& S_{uu} \to \chi \chi \to (u \, A_\chi) (u \, A_\chi) \to (j \, j_{gg}) (j \, j_{gg})     \;\; , \;\;\;  {\rm for} \;\; M_A \ll m_\chi  ~~,
\label{eq:cascade-light}
 \\ [2mm]
&& S_{uu} \to \chi \chi \to  (u \, g) (u \, g) \to (j \, j) (j \, j)     \;\; , \;\;\;  {\rm for} \;\; M_A > m_\chi  ~~.
\label{eq:cascade-right}
\eear
In both mass cases (see diagrams in Figure~\ref{fig:DiagramSchichi}), the cascade decay gives rise to a pair of dijet resonances, each peaked at $m_\chi$. The $4j$ invariant mass distribution of this pair of dijets signal peaks at $M_S$ in a parton level simulation, but parton shower, jet algorithm and detector effects shift it slightly below $M_S$ due to QCD radiation located outside the jet cones.

\begin{figure}[t!]
\hspace*{-0.01cm}  
\includegraphics[width=0.47\textwidth]{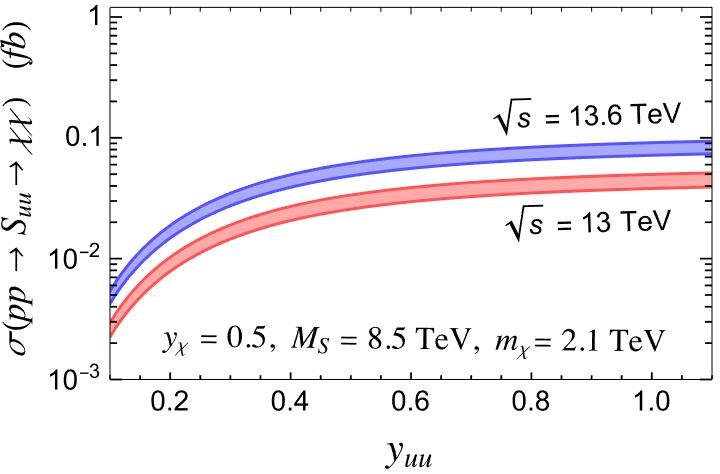}  \hspace*{5mm}  
\includegraphics[width=0.47\textwidth]{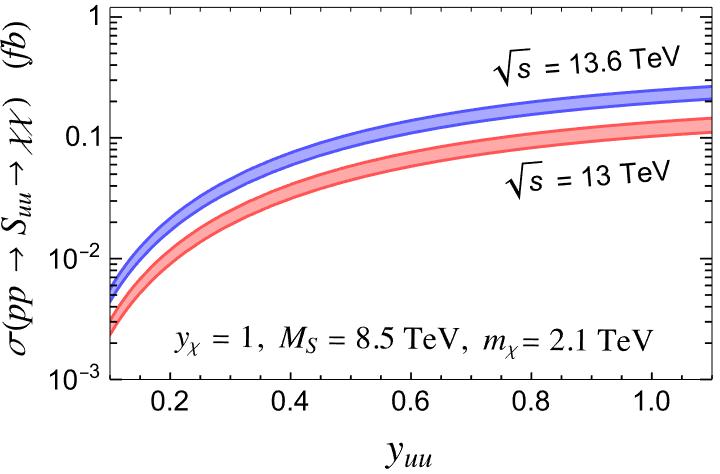}
 \\ [-0.8cm]
\caption{NLO production cross section of the $S_{uu}$ diquark (based on \cite{Dobrescu:2018psr,Han:2009ya}) times the $S_{uu}\to \chi \chi $   branching fraction at the LHC with center-of-mass energy of 13 TeV (lower red bands) or 13.6 TeV (upper blue bands), as a function of  the coupling $y_{uu}$ of $S_{uu}$ to up quarks.
Effective coupling of $S_{uu}$ to $\chi$ vectorlike quarks [see Eq.~(\ref{eq:ychi})] is fixed at $y_\chi = 0.5$ (left panel) or $y_\chi = 1$ (right panel).
The  $S_{uu}$ and $\chi$ masses are fixed at $M_S = 8.5$ TeV, $m_\chi = 2.1$ TeV. 
The band widths represent the PDF uncertainties described in the text.
\\
\vspace*{1mm}
} 
\label{fig:xsecSchi}
\end{figure}

This 4-jet  final state may be the origin of the two striking events observed by CMS \cite{CMS:2022usq} at a $4j$ invariant mass of 8 TeV.
Those events are consistent with $M_S \approx 8.5$ TeV and $m_\chi \approx  2.1$ TeV. The NLO cross section times $S_{uu}$ branching fraction 
is displayed as a function of the $y_{uu}$ coupling  in Figure~\ref{fig:xsecSchi}, for proton-proton collisions with $\sqrt{s} = 13$ TeV and 13.6 TeV.
The left and right plots  refer to particular choices for the $y_\chi$ effective coupling defined in (\ref{eq:ychi}): $y_\chi = 0.5$ and 1, respectively.
This computation used the  NLO PDF sets of MMHT2014 \cite{Harland-Lang:2014zoa} and  CT14   \cite{Dulat:2015mca}; the latter includes larger uncertainties (roughly 12\%), and gives a cross section about 10\% higher than the former.
The widths of the bands in Figure~\ref{fig:xsecSchi} extend from the $1\sigma$ lower limit of  MMHT2014  to the $1\sigma$  upper  limit of  CT14. 

\smallskip

\subsection{Invariant mass distribution of the four jets}
\label{sec:M4j}

\begin{figure}[b!]
\vspace*{2mm}
\hspace*{2.7cm}  \includegraphics[width=9.8cm, angle=0]{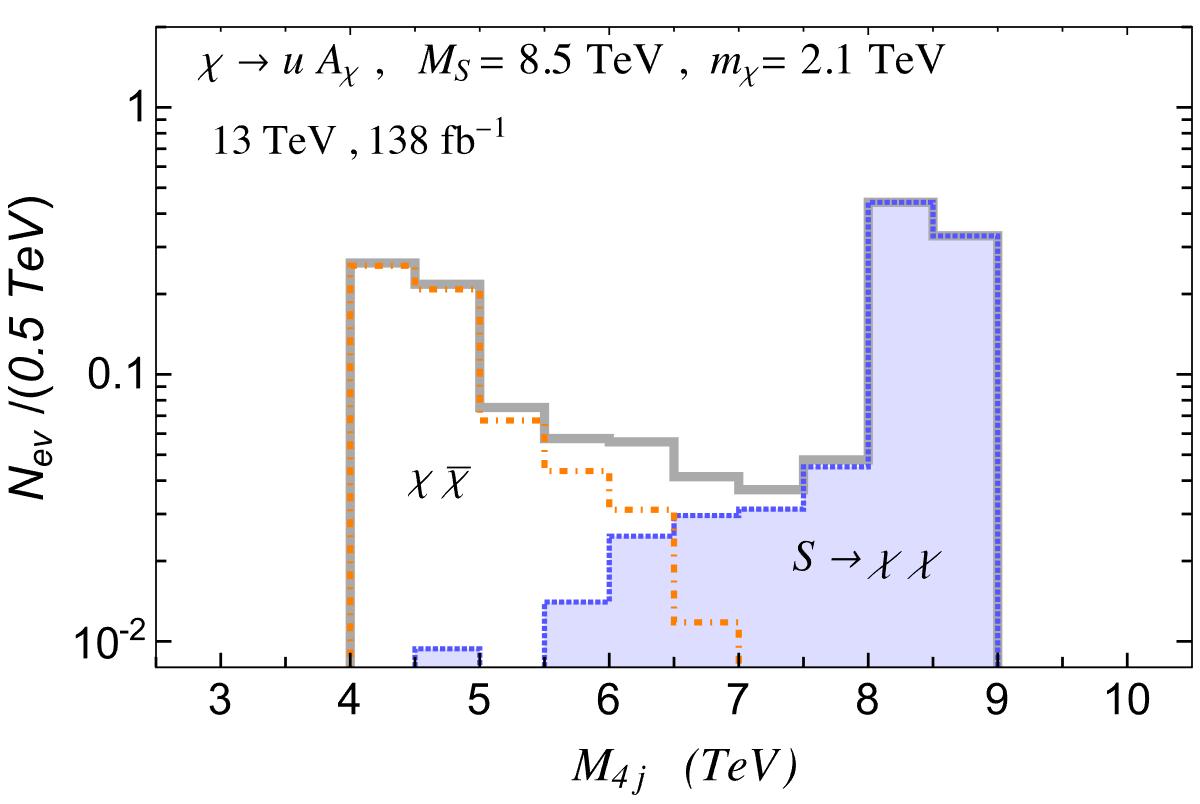}       
\vspace*{-.3cm}  
\caption{The $4j$ invariant mass distribution of the pair of dijets signal simulated at parton level in the $S\chi A$ theory with $S_{uu}$  mass of 8.5 TeV, $\chi$ mass of $2.1$ TeV, and $M_A \ll m_\chi$.
Dashed blue line (which bounds  the shaded region) shows the  number of events due to the resonant process $pp \to S_{uu} \to \chi \chi \to (u \, A_\chi) (u \, A_\chi) \to  4j$ for $y_{uu} = 0.4$, $y_\chi = 0.5$. Dot-dashed orange line shows  the  number of events due to the nonresonant process $pp \to  \chi \overline \chi \to (u \, A_\chi) (\overline  u \, A_\chi) \to 4j$. The solid thick  (gray) line gives the sum of these two distributions.
\vspace*{2mm}
} 
\label{fig:PlotRes4jchi2100}
\end{figure}

In order to estimate the invariant mass distributions, the resonant process (\ref{eq:cascade-light}) with light $A_\chi$   (see the left diagram of Figure~\ref{fig:DiagramSchichi}) is simulated at parton level and leading order with Madgraph \cite{Alwall:2014hca}, using  model files generated with FeynRules \cite{Alloul:2013bka}, and the default PDF set NNPDF23\_lo \cite{Ball:2013hta}.
The acceptance for the event selection described in Section \ref{sec:950GeV} (based on the one used by CMS \cite{CMS:2022usq}) is $A_{4j} \approx 30\%$.
The total $M_{4j}$ distribution is displayed as the solid thick (gray) line in Figure~\ref{fig:PlotRes4jchi2100}. Besides the events due to the resonant process $pp \to S_{uu} \to \chi \chi \to  4j$ (whose distribution is shown by the dashed blue  line for $y_{uu} = 0.4$, $y_\chi = 0.5$), there are unavoidable events due to the nonresonant process $pp \to  \chi \overline \chi \to 4j$ (see the left diagram of Figure~\ref{fig:DiagramsNonreschi}); the $M_{4j}$ distribution of the latter events is shown as the dot-dashed orange line. 

A few other parameters are fixed here, but their impact on the $M_{4j}$ distribution is minor.
The $A_\chi$ mass used in this simulation is $M_A = 10$ GeV; any other choice below 50 GeV or so would not make a significant difference, as the large boost of $A_\chi \to g g $ implies that  the gluons are sufficiently collinear to fit inside a narrow jet cone. 
 Another parameter fixed here is  $y_{\chi\chi_L} = 0$ in (\ref{eq:diquarkChi}), implying
$y_{\chi\chi_R} = y_\chi$, which affects the angular distributions. Finally,  the values of the two  parameters from (\ref{eq:yuk1}) 
are not consequential as long as $10^{-6} \lesssim \lambda_u  \ll 1$ (so that $\chi$ decays promptly, and is a narrow state),
and $10^{-3} \lesssim \lambda_\chi \ll 1$  (so that $A_\chi$ decays promptly, and has suppressed production in gluon fusion).

While the overall normalization of the $M_{4j}$ distribution due to the resonant process is modified for different values of $y_{uu}$ and $y_\chi$ (which affect the $S_{uu}$ production cross section and branching fractions), 
the normalization of the distribution due to the nonresonant process is completely fixed by the QCD coupling.
The sum of these two distributions for $y_{uu} = 0.4$, $y_\chi = 0.5$ corresponds to the solid thick   (gray) line in Figure~\ref{fig:PlotRes4jchi2100}.
For these values of the couplings, $S_{uu}$ is a narrow state with a width-to-mass ratio $\Gamma_S/M_S \approx  0.35\%$, and a production cross section of  $\sigma_S \approx0.043$ fb at $\sqrt{s} = 13$ TeV.

The distinctive double-peak shape of the total $M_{4j}$ distribution was not previously noticed. The simultaneous inclusion of the resonant and the nonresonant contributions is necessary for a correct interpretation of the data, and should be analyzed by the CMS and ATLAS collaborations in future searches of this type.
Based on the parton-level simulation with $y_{uu} = 0.4$, $y_\chi = 0.5$, the expected number of events with $M_{4j} \gtrsim 8$ TeV is approximately 0.8 in 138 fb$^{-1}$ of 13 TeV data, which could easily fluctuate into two events seen at CMS \cite{CMS:2022usq} and no events seen at ATLAS \cite{ATLAS:2023ssk}. 
Once parton shower and realistic detector effects are taken into account, the  total $M_{4j}$ distribution shown in Figure~\ref{fig:PlotRes4jchi2100} would have lower peaks and longer tails, especially towards lower invariant masses.
Note that most nonresonant events with $M_{4j} \lesssim 4.5$ TeV are due to incorrect pairing of the two dijets (this is similar to the case of a lighter $\chi$ shown in  Figure~\ref{fig:plotNonres4j1000}).


\subsection{Other signals: dijet resonance, boosted top quark(s)}
\label{sec:top}

The decay $S_{uu} \to u u $ leads to a dijet resonance signal \cite{Dobrescu:2018psr, Richardson:2011df}. For $y_{uu} = 0.4$, $y_\chi = 0.5$ and the masses considered here, Eqs.~(\ref{eq:uuWidth}) give a dijet branching fraction ${\cal B} (S_{uu} \to uu) \approx 46\%$. Thus, the number of dijet events with a mass near 8.5 TeV 
predicted  in $L_2 = 138$ fb$^{-1}$ of 13 TeV data is $\sigma_S \, {\cal B} (S_{uu} \to uu) L_2 A_{jj} \approx 2.7 A_{jj} $, where $A_{jj} $ is the dijet acceptance, which depends on the $jj$ event selection used by the experimental collaborations (for reviews, see \cite{Harris:2011bh, Dobrescu:2021vak}). Keeping the number of $4j$ events fixed through an adjustment of $y_\chi$, the number of dijet events near 8.5 TeV may be larger if $y_{uu}$ is increased; for example,  $y_{uu} = 0.7$,  $y_\chi = 0.37$ gives 5 times more dijet events than $4j$ events before acceptance is taken into account.
Even though the QCD dijet background is small at a mass as large as 8.5 TeV, it still consists of a few events in 138 fb$^{-1}$ of 13 TeV data \cite{ATLAS:2019fgd, Sirunyan:2019vgj},  and thus probing this $jj$ channel is likely to require more data compared to the $4j$ channel.

The interactions of the $A_\chi$ pseudoscalar assumed so far are given in (\ref{eq:yuk1}).
These may be expanded to include the right-handed top quark:
\be
-  \lambda_t  i A_\chi  \,  \overline \chi_L  \, t_R  + {\rm H.c.}  ~~,
\label{eq:yukt}
\ee
where  $ \lambda_t$ is the new Yukawa coupling, taken here to satisfy $(\lambda_t/\lambda_u)^2 \ll 1$ so that the rates for the processes with jets considered so far are not significantly affected. 
In the Heavy-$A_\chi$ case,  the branching fraction for the 1-loop $\chi$ decay into $t g$ is ${\cal B}(\chi \to t g) \approx (\lambda_t/\lambda_u)^2$, where corrections of order $(m_t/m_\chi)^2$ are neglected.
Thus, the process $p p \to S_{uu} \to \chi \chi \to (t g)(u g)$ (right diagram of Figure~\ref{fig:DiagramSchichi} with one $u$ quark replaced by $t$) has a rate suppressed by a factor of $2 (\lambda_t/\lambda_u)^2$ compared to the $(u g)(u g)$ final state.

\begin{figure}[t!]
\hspace*{2.7cm}  \includegraphics[width=9.7cm, angle=0]{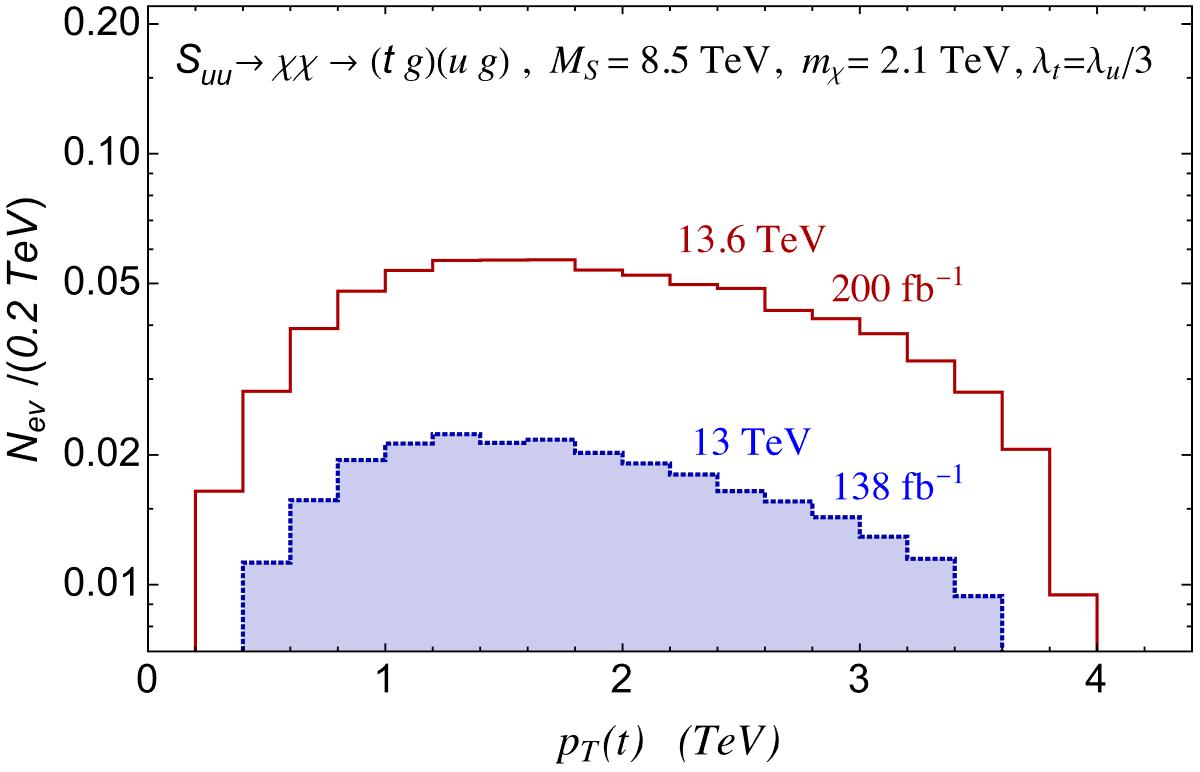}       
\vspace*{-0.2cm}  
\caption{Top quark $p_T$ distribution in the process $pp \to S_{uu} \to \chi \chi \to (t \, g) (u \, g)$ simulated at parton level, using Madgraph \cite{Alwall:2014hca}  at leading order, within the $S\chi A$ theory with a top quark coupling [see (\ref{eq:yukt})] 
$\lambda_t = \lambda_u/3$, for $M_S = 8.5$ TeV, $\chi = 2.1$ TeV, and $M_A > m_\chi$.
Solid red line corresponds to 200 fb$^{-1}$ of 13.6 TeV data.
Dashed blue line (which bounds the shaded region) corresponds to 138 fb$^{-1}$ of 13 TeV data.
\vspace*{1mm}
} 
\label{fig:PlotpTtop.pdf}
\end{figure}

The top quark produced in this process typically has a  high boost. 
Figure~\ref{fig:PlotpTtop.pdf} displays the $p_T$ distribution of the top quark for $\sqrt{s} = 13.6$ TeV (with 200 fb$^{-1}$, solid red line) and 13 TeV (with 138 fb$^{-1}$, dashed blue line). 
In the case of hadronic decays, the highly-boosted top quark would appear in the detector as a narrow jet with a three-prong substructure that may be difficult to disentangle.
In the case of leptonic decays, the lepton has positive electric charge in almost all events of this type, because a $\overline t$ can only be produced in decays originating from  $S_{uu}^\dagger$; the production of this antiparticle requires an $\bar u  \bar u$ initial state, and it has a rate two orders of magnitude smaller than the one for $S_{uu}$.

Even more boosted top quarks may be produced if the ultraheavy $S_{uu}$ couples directly to $\overline t_{R } \,  u^c_{R }$
 or $\overline t_{R } \,  t^c_{R }$, as discussed in \cite{Dobrescu:2019nys}.
In addition, $S_{uu}$ couplings to $\overline t_{R } \,  \chi^c_{R }$ or $\overline u_{R } \,  \chi^c_{R }$ would induce 3-body decays of the vectorlike quark, $\chi \to t u\bar u$ or $\chi \to uu\bar u$ (through an off-shell $S_{uu}$), leading to final states of the type  $tt + 4j$ or $6j$. 

\section{$S2\chi A$ theory: resonant and nonresonant dijet pairs} 
\setcounter{equation}{0} \label{sec:2Vquarks}

This Section focuses on a new renormalizable theory, referred to as $S2\chi A$, which combines relevant features of the theories presented in Sections 2 and 3.
Its implications are novel signatures at the LHC, and simultaneous explanations for the 3.9$\sigma$ excess at 8 TeV and the 3.6$\sigma$ excess at 0.95 TeV reported by CMS \cite{CMS:2022usq}.

In addition to the SM, the $S2\chi A$ theory includes four fields: a scalar diquark $S_{uu}$ of SM gauge charges (6, 1, +4/3),
two vectorlike quarks $\chi_1$  and $\chi_2$ of charges (3, 1, +2/3), and a gauge-singlet pseudoscalar $A_\chi$. Besides SM gauge interactions, the interactions involving the four new fields are given by the following terms in the Lagrangian:
\be \hspace*{-0.2cm}
 S_{uu}  \! \left(   \frac{y_{uu}}{2} \,  \overline u_{ _R}  u^c_{ _R}
+ \frac{y_{\chi}}{2}    \, \overline \chi_{2_R}  \chi^c_{2_R }   
 \! \right)  
- i A_\chi \! \left( \rule{0mm}{4.2mm} \right. \!\!  \lambda_{_{21}}   \overline \chi_{2_L} \chi_{1_R}  + \! \sum_{\ell = 1,2} \! \lambda_{\ell u}  \overline \chi_{\ell_L} u_{_R}  
+  \lambda_{_{11}}  \overline \chi_{1_L}  \chi_{1_R}   \!\!
\left.\rule{0mm}{4.2mm} \right) 
   + {\rm H.c.} 
  \label{eq:chi2chi1}
\ee
Here $y_{uu},  y_{\chi},  \lambda_{_{\ell 1}},  \lambda_{\ell  u}$ ($\ell = 1,2$) are real parameters. 
The above Lagrangian terms are written in the mass eigenstate basis, where the mass terms for the vectorlike quarks are 
$-  m_{\chi_\ell}   \overline \chi_{\ell}  \chi_{\ell}$. 
Several other possible interactions, such as 
$S_{uu} \, \overline  \chi_{2_R}  \chi^c_{1_R } $ or  $A_\chi  \overline \chi_{1_L} t_{_R}$, are ignored here for simplicity. 

In what follows only the case $M_A \ll m_{\chi_1}$ is considered (so that $A_\chi \to gg$ behaves as a single narrow jet $j_{gg}$), but similar results can be obtained for $M_A > m_{\chi_2}$ (which leads at one loop to $\chi_2 \to \chi_1 g$ and $\chi_\ell \to u g$, following the arguments of  Section \ref{sec:heavy}).
As discussed in Section \ref{sec:1tev}, various constraints require $  \lambda_{\ell 1} ,  \lambda_{\ell u}  \ll 1$ when $A_\chi $ has a mass in the $5-50$ GeV range. 
The $\lambda_{1 1}$ coupling induces the 1-loop decay of $A_\chi $ into two gluons, 
while the $\lambda_{1 u}$ coupling induces the tree-level decay $\chi_1 \to u A_\chi$. These decays have branching fractions of nearly 100\% independent of the values of those two couplings, and they are prompt as long as $\lambda_{1 1}$  and $\lambda_{1 u}$ are not too small (see Section  \ref{sec:light}).
The other two couplings of $A_\chi$ in (\ref{eq:chi2chi1}), $\lambda_{2 u}$ and $\lambda_{2 1}$, trigger the $\chi_2$ decays.
There are two tree-level decay modes: $\chi_2 \to u \, A_\chi$ and $\chi_2 \to \chi_1 \, A_\chi$. Their leading-order widths are given by 
\bear
 && \Gamma (\chi_2 \to u \, A_\chi) =  \dfrac{  \lambda_{2 u}^2 }{32 \pi}  \, m_{\chi_2} ~~,  \nonumber
 \\ [-1mm]
 && 
 \\ [-2mm]
 && \Gamma (\chi_2 \to \chi_1 \, A_\chi)  =   \dfrac{ \lambda_{21}^2 }{32 \pi} \, m_{\chi_2}   \left( 1 - \dfrac{m_{\chi_1}^4 }{m_{\chi_2}^4 } \right) ~~.  \nonumber
\label{eq:uuWidth}
\eear
The $u u \!\to\! S_{uu} \!\to\! \chi_2\chi_2$ production proceeds as discussed in Section \ref{sec:minimal}, with cross section shown in Figure~\ref{fig:xsecSchi}.
When both $\chi_2$ quarks decay to $u  A_\chi$, the final state is $(u A_\chi) (u A_\chi)  \to 4j$ as shown in the  left diagram of Figure~\ref{fig:DiagramSchichi} with $\chi$ replaced by $\chi_2$.
The rate for this process is lower than in Figure~\ref{fig:PlotRes4jchi2100} due to an extra branching  factor of ${\cal B} (\chi_2 \to u \, A_\chi)^2$. 

\begin{figure}[t!]
\hspace*{-0.1cm}  %
\includegraphics[width=0.49\textwidth]{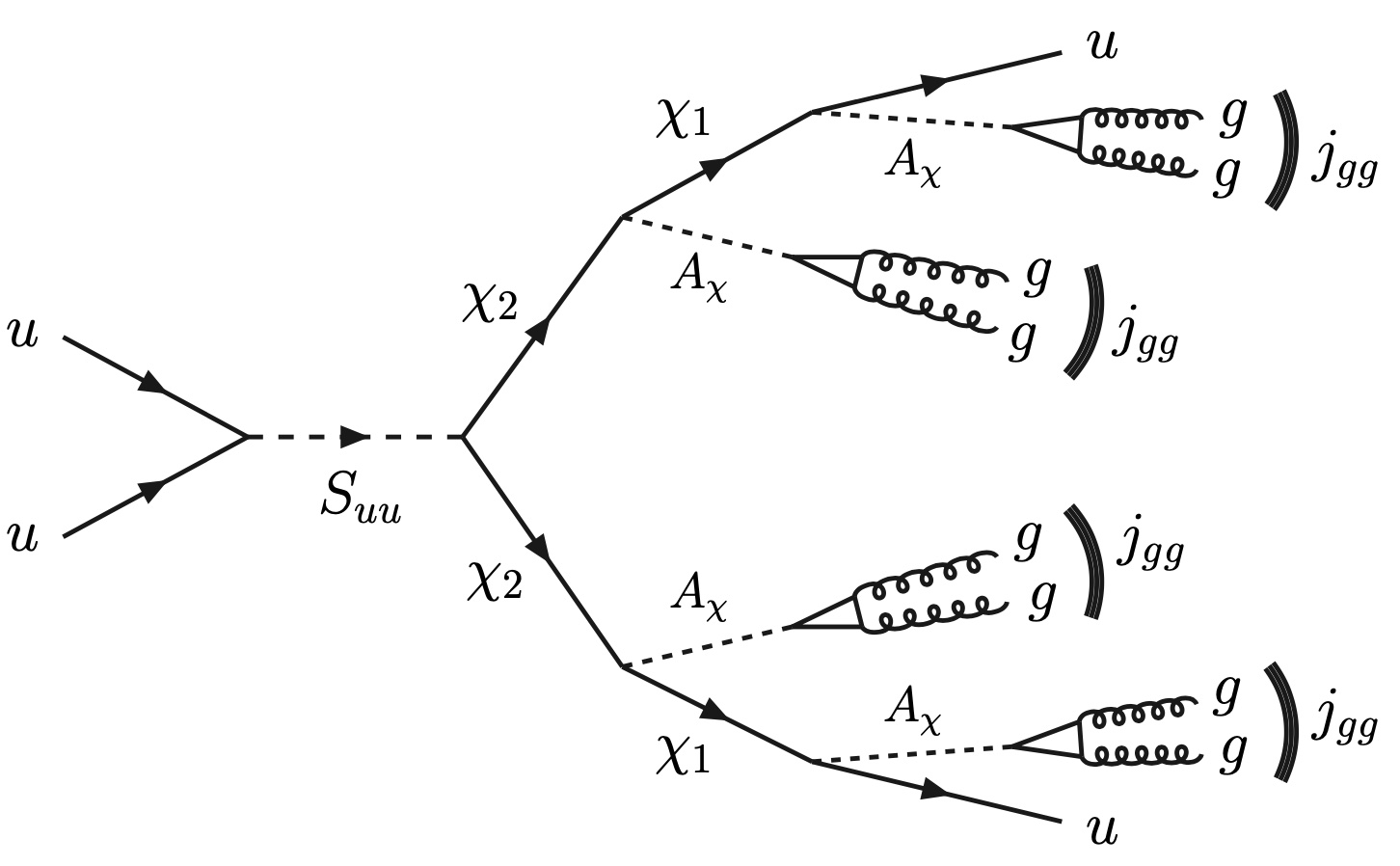}    \\ [-4.9cm]
\hspace*{8.17cm}  \includegraphics[width=0.49\textwidth]{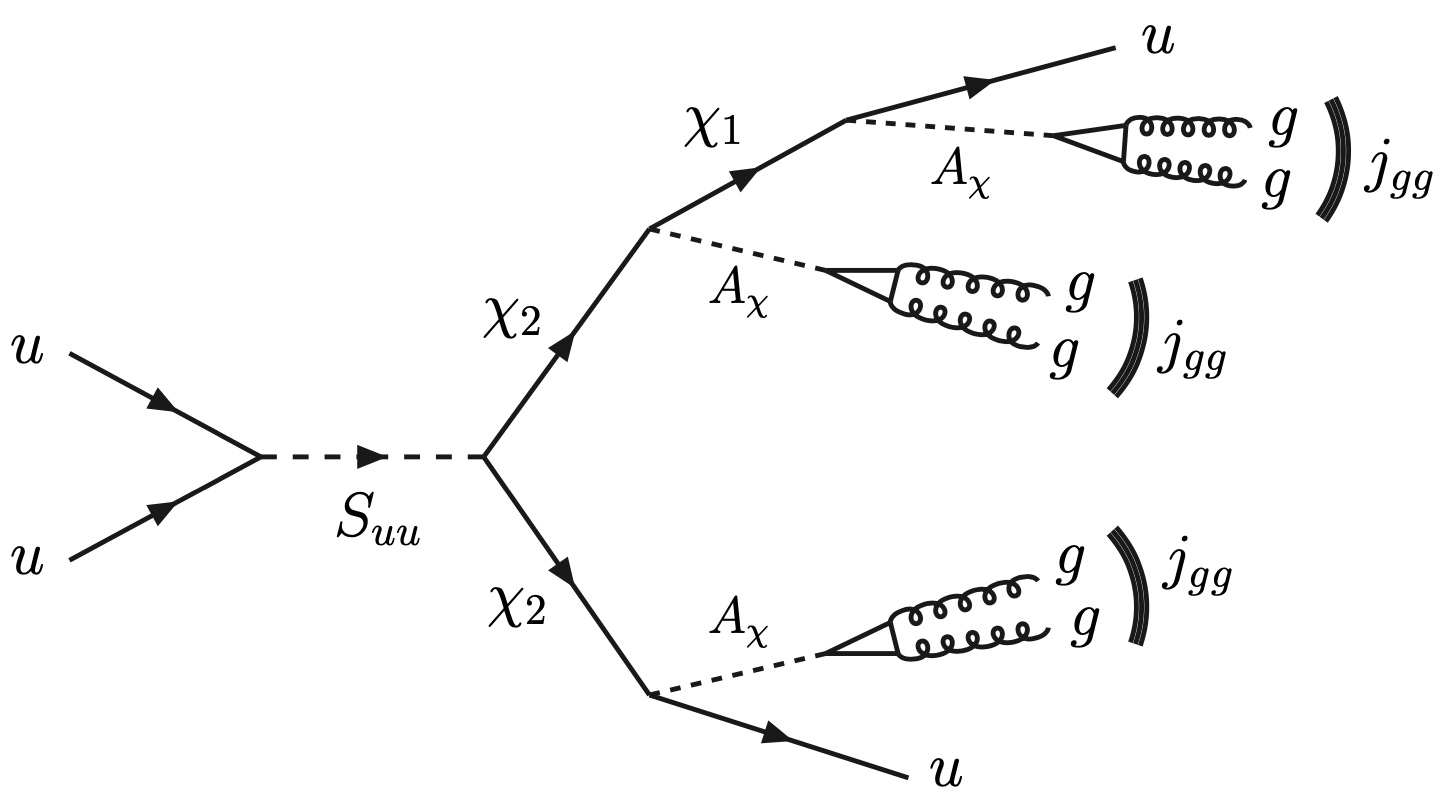}     
 \\ [-0.2cm]
\caption{LHC $s$-channel production of the $S_{uu}$ diquark  followed by a cascade decay
via $\chi_2\chi_2$ in the $S2\chi A$ theory for $M_A \ll m_{\chi_1}$. Left diagram: 
each $\chi_2$  decays into $\chi_1 \,  A_\chi$, leading to the final state $4A_\chi + uu \to 6 j$.
Right diagram:   one $\chi_2$  decays into $\chi_1 A_\chi$ and the other one  decays into $u A_\chi$, leading to the final state $3A_\chi + uu \to 5 j$.
Analogous diagrams, with $A_\chi$ replaced by a single gluon (coupled at one loop), exist in the $M_A > m_{\chi_1}$ case.
\vspace*{3mm}
} 
\label{fig:DiagramSchichi56}
\end{figure}

There are however two other processes that contribute to final states with dijet pairs. Although these final states have additional signal jets,
the existing searches for pairs of dijet resonances, performed by CMS \cite{CMS:2022usq} and ATLAS \cite{ATLAS:2023ssk}, select the  four jets of highest-$p_T$  in each event and pair them independently of lower-$p_T$ jets contained in the same event.
When both $\chi_2$ quarks decay to $\chi_1  A_\chi$, the $u u \to S_{uu} \to \chi_2\chi_2$ production leads to the final state $4A_\chi + uu \to 6 j$  displayed in the left diagram of Figure~\ref{fig:DiagramSchichi56}. The rate in that case has an extra branching factor of  ${\cal B} (\chi_2 \to \chi_1 \, A_\chi)^2$. 
There is also the mixed process, where one $\chi_2$ decays to $u  A_\chi$ and the other one decays to $\chi_1  A_\chi$, with a final state $3A_\chi + uu \to 5j$ (right diagram of Figure~\ref{fig:DiagramSchichi56}) and an extra branching factor of  $2 {\cal B} (\chi_2 \to u \, A_\chi) \,  {\cal B} (\chi_2 \to \chi_1 \, A_\chi)$.

Besides the three resonant processes (originating from an $s$-channel $S_{uu}$) discussed above, there are contributions to the dijet pairs signal from  nonresonant processes initiated by the interactions of the vectorlike quarks with gluons. The $pp \to \chi_1   \overline{\chi}_1 \to 4j$ process has the highest rate because $ \chi_1$ is lighter, but its contributions are at lower $M_{4j}$ where the QCD background is large. 
There are also processes initiated by  $pp \to \chi_2   \overline{\chi}_2$ which lead to the $(u A_\chi)(\overline u A_\chi)$,  $3A_\chi + u \overline u$,  $4A_\chi + u\overline u$ final states.

The LHC phenomenology of the $S2\chi A$ theory depends mostly on three masses ($M_S$, $m_{\chi_2}$, $m_{\chi_1}$) and three dimensionless parameters ($y_{uu}$, $y_\chi$, $\lambda_{2u}/\lambda_{21}$). The excess events reported by CMS (in both resonant and nonresonant dijet pairs) can be simultaneously explained if the masses are fixed at 
\be
M_S \approx 8.5  \; {\rm TeV}  \;\;   , \;\;\;   m_{\chi_2} \approx 2.1  \; {\rm TeV}  \;\;   , \;\;\;  m_{\chi_1} \approx 0.95   \; {\rm TeV}   ~~.
\label{eq:masspoint}
\ee
The $y_{uu}$ coupling controls the $S_{uu}$ production cross section, and the ratio $y_\chi / y_{uu}$ determines the $S_{uu}$ branching fractions. 
The ratio $\lambda_{2u}/\lambda_{21}$ determines the $\chi_2$ branching fractions. For concreteness, the benchmark choice for these dimensionless parameters 
\be
y_{uu} = 0.8  \;\;   , \;\;\;   y_\chi = 1 \;\;   , \;\;\;  \lambda_{2u}/\lambda_{21}  = 0.7 ~~
\label{eq:bench}
\ee
gives the following production cross sections at $\sqrt{s} = 13$ TeV:
$\sigma (p p \to S_{uu} ) \approx  0.17$ fb for the resonant process, as well as  $\sigma (p p \to   \chi_2 \overline  \chi_2) \approx  6.7 \times 10^{-2}$ fb and 
$\sigma (p p \to   \chi_1 \overline  \chi_1) \approx  54$ fb for the two nonresonant productions. The $S_{uu}$  width-to-mass ratio is $\Gamma_S/M_S =  1.4\%$, so $S_{uu}$ is a narrow resonance.
The relevant branching fractions of $S_{uu} $ and $\chi_2$ are given for the above benchmark by 
$ {\cal B} (S_{uu}  \to \chi_2 \, \chi_2) \approx 54\%$ and $ {\cal B} (\chi_2 \to \chi_1 \, A_\chi)\approx 66\%$.

\begin{figure}[t!]
\hspace*{2.1cm}  \includegraphics[width=10.4cm, angle=0]{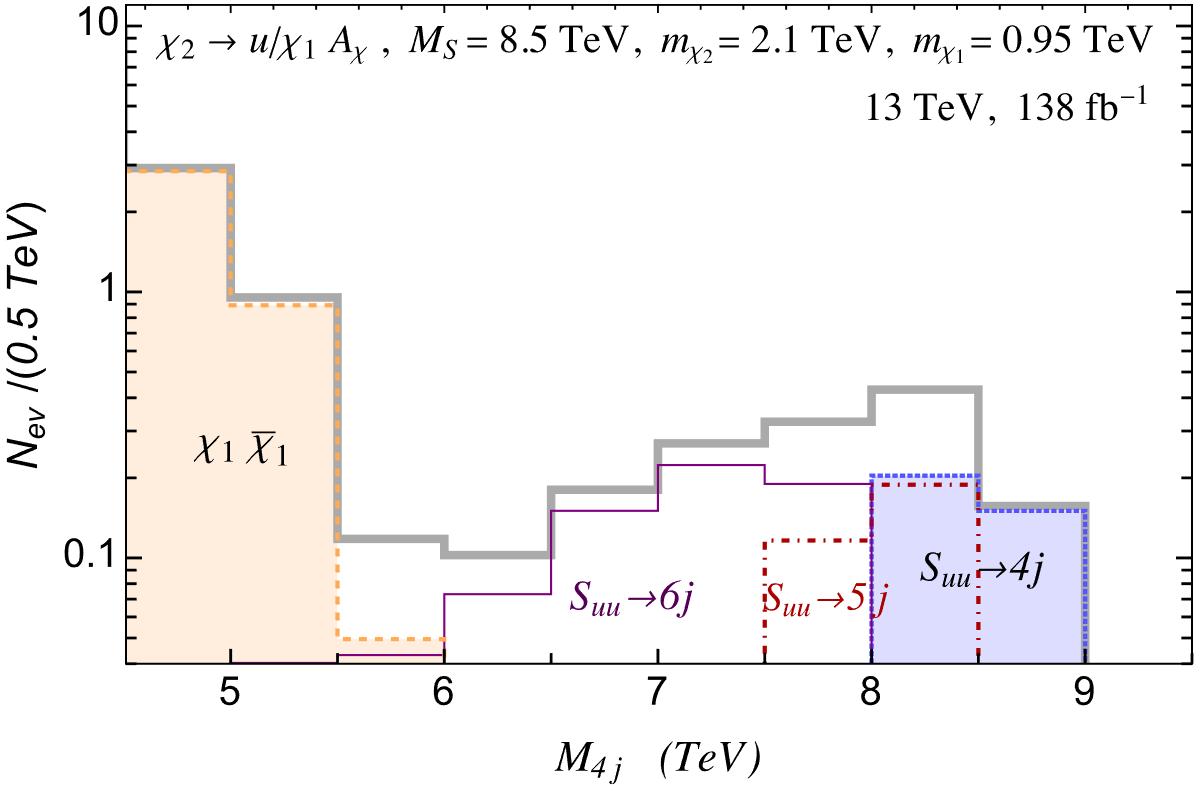}      
\vspace*{-.4cm}  
\caption{The $4j$ invariant mass distribution of the dijet pair signal simulated at parton level in the $S2 \chi  A$ theory with masses (\ref{eq:masspoint}) and couplings (\ref{eq:bench}), corresponding to a narrow  $S_{uu}$.
Dashed blue line (boundary of the right shaded region) represents the number of events due to $pp \!\to\! S_{uu} \!\to\!  \chi_2 \chi_2 \!\to\!  (u \, A_\chi) (u \, A_\chi)$,  while dot-dashed red line and solid purple line show the invariant mass distributions of the four leading jets  from $S_{uu} \!\to\!  \chi_2 \chi_2 \!\to\!  (\chi_1 \, A_\chi) (u/\chi_1 \, A_\chi) \!\to\!   5j/6j $, respectively. Dashed orange line  (boundary of  the left shaded region) shows number of events due to the nonresonant process $pp \!\to\!  \chi_1 \overline \chi_1 \!\to\! (u \, A_\chi) (\overline  u \, A_\chi) \!\to\! 4j$. 
Thick solid (gray) line is the sum of all these distributions.
\vspace*{1.7mm}
} 
\label{fig:M4jPlotRes2chi}
\end{figure}

The total invariant mass distribution of the four leading jets, $M_{4j}$, obtained from a parton-level and leading-order Madgraph simulation (see Section~\ref{sec:M4j}) with parameters given in (\ref{eq:masspoint}) and (\ref{eq:bench}),  is displayed as the solid thick (gray) line in Figure~\ref{fig:M4jPlotRes2chi}, for $\sqrt{s} = 13$ TeV and integrated luminosity of 138 fb$^{-1}$.
 Also shown there are the resonant $(u A_\chi)(u A_\chi)$ $M_{4j}$ distribution (dashed blue line boundary of the right shaded region) with a peak at $M_S$, 
the $M_{4j}$ distributions of the four leading jets due to the $S_{uu} \to 3A_\chi + uu \to 5j$ (dot-dashed red line) and $S_{uu} \to 4A_\chi + uu \to 6j$ (solid purple line) processes,
and the $M_{4j}$ distribution due to nonresonant  $\chi_1   \overline{\chi}_1$ production (dashed orange line boundary of the left shaded region). 
Although the latter has by far the largest rate, it is concentrated mostly at  $M_{4j} < 4.5 $ TeV (see also Figure~\ref{fig:plotNonres4j1000}), a region not included in Figure~\ref{fig:M4jPlotRes2chi} because of the large QCD background.
The contribution from nonresonant  $\chi_2  \overline{\chi}_2$ production is also included in the  total $M_{4j}$, but it is too small to appear separately in the plot.

\begin{figure}[h!]
\hspace*{2.1cm}  \includegraphics[width=10.2cm, angle=0]{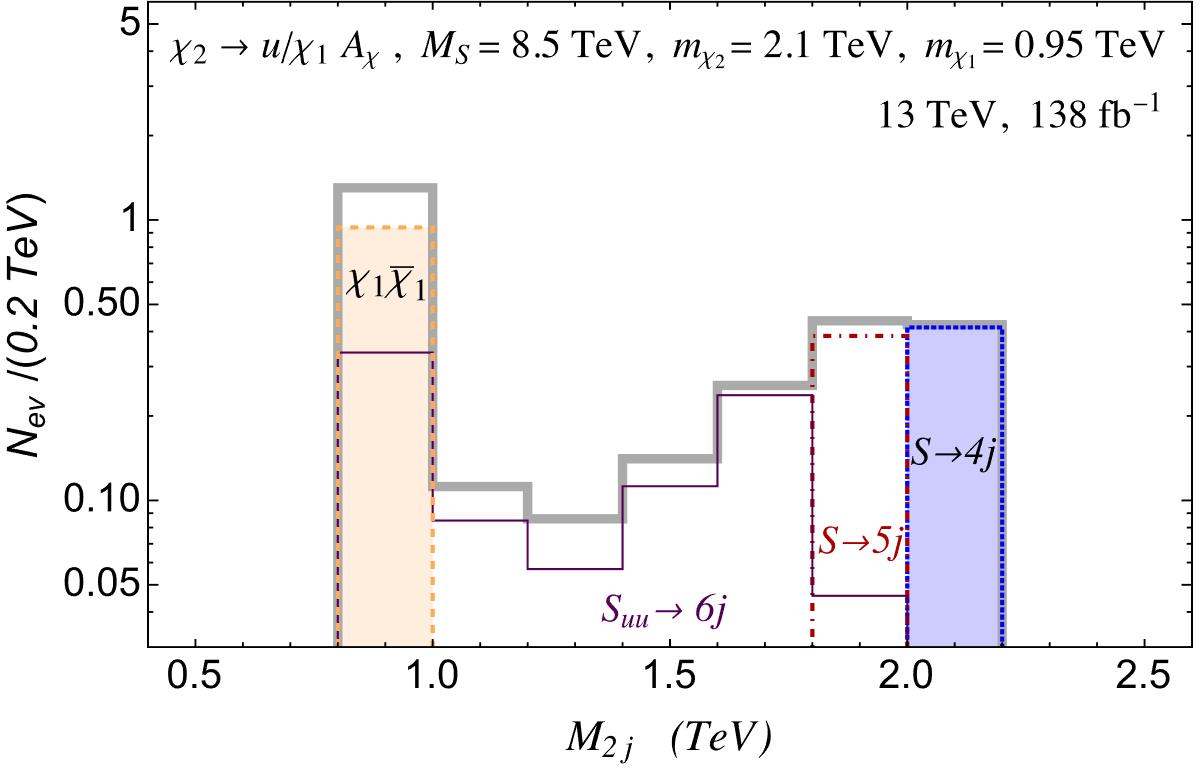}      
\vspace*{-.4cm}  
\caption{Average dijet invariant mass distribution ($M_{2j}$)
of the dijet pair signal simulated at parton level  for the same processes and parameters as in Fig.~\ref{fig:M4jPlotRes2chi} (only events with  $M_{4j} > 5$ TeV are included).
\vspace*{0mm}
} 
\label{fig:M2jPlotRes2chi}
\end{figure}

It is interesting to compare the $M_{4j}$ distribution of this $S2 \chi  A$ theory (Figure~\ref{fig:M4jPlotRes2chi}) with that obtained in Section~\ref{sec:minimal} for  the $S \chi  A$ theory  (Figure~\ref{fig:PlotRes4jchi2100}).
Both of them have a double-peak shape, but $S2 \chi  A$ predicts a larger ratio between the 
number of events with 5 TeV $ < M_{4j} < 8$ TeV and  $ M_{4j} > 8$ TeV.
The increased population of events with $M_{4j}$ in the   $5-8$ TeV range is due to resonant production of events with 5 or 6 jets, and also due to nonresonant $\chi_1  \overline{\chi}_1$ production.
To understand better that population, Figure~\ref{fig:M2jPlotRes2chi} shows the
average dijet invariant mass distribution ($M_{2j}$)
of the dijet pair signal simulated  at parton level  for the same processes and parameters as in Figure~\ref{fig:M4jPlotRes2chi}, with the additional 
constraint that only events with  $M_{4j} > 5$ TeV are included. It is clear that the $S_{uu} \to 3A_\chi + uu \to 5j$ process (dot-dashed red line in both plots) produces events with $M_{4j} > 7.5$ TeV and $M_{2j}$ near 2 TeV, while the events due to the  $S_{uu} \to 4A_\chi + uu \to 6j$ process (solid purple lines) have smaller $M_{4j}$ and $M_{2j}$. 

Overall, the  $S2 \chi  A$ theory predicts events with $M_{2j}$  clustered in two regions: in the $1.8-2.2$ TeV range, or near 1 TeV.
Once parton shower, hadronization, a jet algorithm, and especially realistic detector effects are included, the  $M_{4j}$ and  $M_{2j}$ distributions will 
have broader and smaller peaks compared to Figures~\ref{fig:M4jPlotRes2chi} and \ref{fig:M2jPlotRes2chi}.  
Furthermore, NLO QCD and parton shower effects could lead to some events in which additional gluons may produce a jet with $p_T$ large enough to   become one of the leading four jets, and thus to affect the pairing selection.  
Nonetheless, dedicated searches for the processes predicted in this theory would provide valuable information.

At $\sqrt{s} = 13.6$ TeV, the production cross sections increase by factors of 1.86 for  $\sigma (p p \to S_{uu} ) $,
1.22 for the nonresonant $\sigma (p p \to   \chi_1 \overline  \chi_1)$ process, and 1.45 for $\sigma (p p \to   \chi_2 \overline  \chi_2)$, given the three masses in (\ref{eq:masspoint}). Thus, from Run 2 to Run 3 of the LHC, the size of the left shaded region in Figure~\ref{fig:M2jPlotRes2chi} increases slower than the other components of the $M_{2j}$  distribution.

Besides signals involving multiple jets, the $S2 \chi  A$ theory may include LHC signatures involving highly boosted top quarks, provided there are some interactions beyond those included in (\ref{eq:chi2chi1}). Various signals of this type are studied in Section~\ref{sec:top} and also in \cite{Dobrescu:2019nys},
but the presence of the not-so-heavy vectorlike quark $\chi_1$ opens further possibilities.
For example,  an $A_\chi \overline \chi_{1_L} t_{_R}$ interaction would induce a nonresonant $ \chi_1 \overline \chi_1 \to t +3j$ process with a single top quark that is less boosted compared to those discussed in Section~\ref{sec:top}

\bigskip

\section{Conclusions} 
 \label{sec:conclusions}    \setcounter{equation}{0} 

As demonstrated in Section~\ref{sec:1tev}, a vectorlike quark $\chi$ coupled to a gauge-singlet pseudoscalar $A_\chi$ can predominantly decay into two hadronic jets for two ranges of their mass ratio $M_A/m_\chi$. A loop process, with $A_\chi$  and $\chi$ running in the loop, induces the $\chi \to u g$ decay when $M_A/m_\chi > 1$. 
For $M_A/m_\chi \ll 1$, the tree-level decay $\chi \to u A_\chi$ followed by the loop decay $A_\chi \to gg$ appears in the detector as a $\chi \to jj$ process, because the gluons arising from a highly-boosted  $A_\chi$  are nearly collinear and form a single narrow jet.

Either of these dijet decays may explain the $3.6\sigma$ excess observed by CMS \cite{CMS:2022usq} in the search for nonresonant production of a pair of dijet resonances at $M_{2j} \approx 0.95$ TeV. The cross section times branching fraction times acceptance computed in Section~\ref{sec:950GeV} for 
$m_\chi = 0.95$ TeV fits well the CMS excess even though there are no adjustable parameters ($\chi \overline \chi$ production is controlled only by the QCD coupling).

The renormalizable theory $S2\chi A$ introduced in Section~\ref{sec:2Vquarks}, involving a diquark $S_{uu}$, two vectorlike quarks $\chi_1$, $\chi_2$,  and $A_\chi$, may account not only for the above nonresonant excess, but also for the two striking $4j$ events with  $M_{4j} \approx 8$ TeV reported by CMS \cite{CMS:2022usq}  (a  $3.9\sigma$ local excess over the SM background). Furthermore, the longer cascade decays of $S_{uu}$  predicted by $S2\chi A$ lead to $5j$ and $6j$ events that would appear in the dijet pair search as having $M_{4j}$ in the $5-8$ TeV range, and $M_{2j}$ clustered around 1 TeV and also around 2 TeV;  this may be the origin of the outlier event with  $M_{4j} \approx 6.6$ TeV observed by ATLAS \cite{ATLAS:2023ssk}.

The NLO cross section for $S_{uu}$ production with $M_S = 8.5$ TeV at the 13.6 TeV LHC increases by a factor of 1.86  compared to the 13 TeV LHC \cite{Dobrescu:2018psr}.
The corresponding increase in the QCD production of $\chi_1 \overline\chi_1 $ with $m_{\chi_1} = 0.95$ TeV is by a factor of 1.22.
Thus, the hypotheses discussed here will be conclusively tested with roughly 300 fb$^{-1}$ of Run 3 data. 

In addition to the above searches for dijet pairs, there are various other final states that ATLAS and CMS can pursue, probing the properties of these heavy colored particles. For example, final states that include highly-boosted top quarks (and no top antiquarks) are expected, as discussed in Section~\ref{sec:top}.  
Independently of the current excess events, further studies of diquarks, which are the only particles that may open a direct window into the 10 TeV scale at the LHC \cite{Dobrescu:2019nys}, are warranted. 

\medskip
\bigskip\bigskip\bigskip


{\bf Acknowledgments:} \  I would also like to thank Robert Harris, Niki Saoulidou, Ilias Zisopoulos, Eva Haldakiakis, Elias Bernreuther, Patrick Fox,  Max Knobbe, Stefan H\"{o}che, Calin Alexa, Ioan Dinu, Ioana Duminica, and Adam Jinaru, for insightful comments and discussions.  
I am grateful to the NCSR Demokritos Institute and NKUA for hospitality during final stages of this work, in the context of the H.F.R.I research grant ``Discovery Hubs with Jets at CMS"/HFRI-20996.
This work was supported by Fermi Research Alliance, LLC, under Contract No. DE-AC02-07CH11359 with the U.S.
Department of Energy, Office of Science, Office of High Energy Physics. 

\smallskip


\end{document}